\documentclass[%
 reprint,
 superscriptaddress,
 amsmath,amssymb,
 prb,
 aps,
floatfix,
]{revtex4-2}

\usepackage{graphicx}
\usepackage{dcolumn}
\usepackage{bm}
\usepackage{amsthm}
\usepackage{upgreek}
\usepackage{pifont}

\usepackage[dvipsnames]{xcolor}
\usepackage{tikz}

\begin{document}

\preprint{APS/123-QED}

\title{Real-time dynamics of 1D and 2D bosonic  quantum matter deep in the many-body localized phase}

\author{Sun Woo Kim} \email{sunwoo-kim@outlook.com}
\affiliation{Max-Planck-Institut f\"ur Physik Komplexer Systeme, N\"othnitzer Stra{\ss}e 38,  01187-Dresden, Germany}
\author{Giuseppe De Tomasi}
\affiliation{Cavendish Laboratory, JJ Thomson Avenue, Cambridge,
CB3 0HE, United Kingdom}
\author{Markus Heyl}
\affiliation{Max-Planck-Institut f\"ur Physik Komplexer Systeme, N\"othnitzer Stra{\ss}e 38,  01187-Dresden, Germany}

\date{\today}

\begin{abstract}
Recent experiments in quantum simulators have provided evidence for the Many-Body Localized (MBL) phase in 1D and 2D bosonic quantum matter. The theoretical study of such bosonic MBL, however, is a daunting task due to the unbounded nature of its Hilbert space. In this work, we introduce a method to compute the long-time real-time evolution of 1D and 2D bosonic systems in an MBL phase at strong disorder and weak interactions. We focus on local dynamical indicators that are able to distinguish an MBL phase from an Anderson localized one. In particular, we consider the temporal fluctuations of local observables, the spatiotemporal behavior of two-time correlators and Out-Of-Time-Correlators (OTOCs). We show that these few-body observables can be computed with a computational effort that depends only polynomially on system size but is independent of the target time, by extending a recently proposed numerical method [Phys. Rev. B \textbf{99}, 241114 (2019)] to mixed states and bosons. Our method also allows us to surrogate our numerical study with analytical considerations of the time-dependent behavior of the studied quantities. 
\end{abstract}

\maketitle

\section{Introduction}
Many-body localization (MBL) generalizes the concept of Anderson Localization (AL) \cite{anderson1958absence} to the interacting regime and has emerged as a novel paradigm for ergodicity breaking of generic many-body systems subject to strong disorder \cite{Abanin_review_2019, Rahul_2015, ALET2018498, BASKO20061126}. 

Experimentally, evidence for the MBL phase has been provided by the use of synthetic quantum platforms based on cold-atoms and trapped ions~\cite{Choi1547, Smith2016, Xu_2018, Guo_2020, Schreiber842, Bordia_2017, L_schen_2017} also covering the case of bosons in two dimensions in Ref. \onlinecite{Choi1547}. However, on the theoretical side, apart from a few numerical studies for small systems \cite{wahl2019signatures, chen2021many}, most of the works have been confined to fermionic/spin models \cite{Prosen_08, Pal_2010, Jonas_2014, Bera_2015, Luitz_2015,DeTo_2017, De_Tomasi_frag_2019, Shreya_2017, Abanin_cri_2015,Serbyn_poer_2016,Khemani_2017}. Indeed, the study of bosonic system out-of-equilibrium is particularly challenging due to its unbounded Hilbert space.

\begin{figure}[b]
\includegraphics[width=1.0\columnwidth]{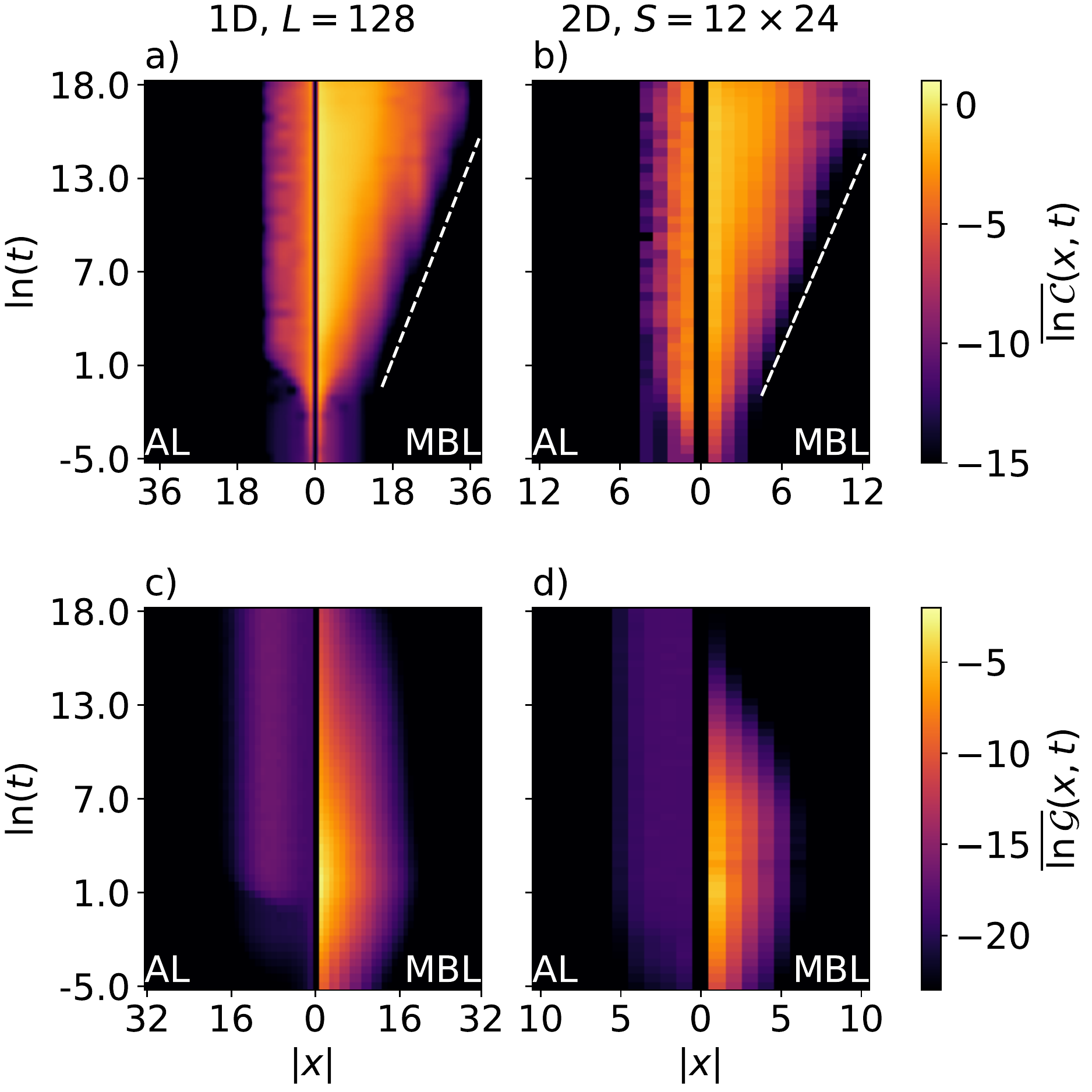}
\caption{\label{fig:colorplots} Disorder-averaged color plot of (a), (b): Out-of-time-ordered commutator $\overline{\ln \mathcal{C}}(x,t)$, and (c), (d): Two-time correlator $\overline{\ln \mathcal{G}}(x,t)$ for (a), (c): 1D and (b), (d): 2D disordered Bose-Hubbard model.}
\end{figure}

In this work, we study the out-of-equilibrium quantum dynamics of bosonic systems deep in an MBL phase in one (1D) and two dimensions (2D). We formulate a method to compute the MBL dynamics of bosons in a controlled approximate fashion by extending a recently proposed method of Ref. \onlinecite{detomasi19} to mixed states and bosonic systems. In particular, we focus on \textit{local dynamical indicators} which are able to distinguish an AL phase from an MBL phase. These dynamical indicators range from temporal  fluctuations of local observables to two-point correlation functions, some of which may be used in a cold atom experimental setup, since they involve only few-point correlation functions \cite{DeTo_2017,Quantum_serby_2014, Inter_Serbyn_2014,Roy_2015}, and 
out-of-time-ordered correlators (OTOCs) describing information scrambling in quantum many-body systems \cite{swingle2017,larkin1969quasiclassical}. We find that bosonic MBL systems exhibit a logarithmic light-cone for information propagation, see Fig.~\ref{fig:colorplots}~(a,b) for the OTOC in 1D and 2D, respectively.

In an MBL phase, particles, despite their interactions, are localized in space, which  imposes strong constraints on the dynamics of the system. As a consequence, particle and energy transport are absent and the system retains local information about its past even for asymptotically long times, as opposed to conventional thermalizing systems \cite{Prosen_08, Pal_2010, Jonas_2014, Bera_2015, Luitz_2015,DeTo_2017}. However, for fermionic and spin systems it is by now understood that interactions induce a dephasing  mechanism, which allows entanglement and quantum correlations to spread during the dynamics, even though transport is absent \cite{Bar_2012, Abanin_2013, Prosen_08}. The entanglement spreading combined with the absence of transport is best understood in terms of an emergent form of integrability and the existence of an extensive number of (quasi-) local integrals of motion (LIOMs) which are coupled by exponentially decaying interactions \cite{Serbyn_con_2013, Huse_phen_2014, Imbrie2016, Ros2015}.
As a result, the time evolution produces dephasing of the LIOMs and therefore a logarithmically slow spread of entanglement. On the other hand, due to the local nature of the LIOMs, transport is strongly hindered and charge relaxation is forbidden \cite{Bar_2012, Abanin_2013, Huse_phen_2014, detomasi19}. In this work we show that this dephasing mechanism also applies to bosonic MBL matter leading to similar phenomenology as has been found for fermions and spins.

The paper is structured as follows. In Sec. \ref{sec:methods}, we describe the method in detail. After detailing the class of models considered, in Sec. \ref{sec:lbitphenomenology}, we describe the $\ell$-bit phenomenology of MBL systems, after which we show how to construct an effective $\ell$-bit Hamiltonian in Sec. \ref{subsec:constructlbit}. Then, in Sec. \ref{subsec:effcalcobs} we discuss methods to compute few-body observables, noting that special Gaussian initial states can be chosen in order to efficiently compute quantities in the case of bosons. We also discuss the the range of validity of our method in Sec. \ref{subsec:rangeofvalidity}. Then, in Sec. \ref{sec:bosonicmbl}, we apply the method to study the disordered Bose-Hubbard model focusing on the following local dynamical indicators. We first study the temporal fluctuations in Sec. \ref{subsec:temporalflucs}. Then, we discuss how the two-time commutator is an indicator for quantum information spreading, and study its fluctuation, a two-point correlator, in  Sec. \ref{subsec:fluctwotime}, as well as its second moment, the OTOC, in Sec. \ref{subsec:otocs}. Finally in Sec. \ref{subsec:occamp}, we study the dependency of these observables on the occupation amplitude, a property unique to bosonic systems.

\section{Methods} \label{sec:methods}
In this Section, we describe the model and the method used through the work. A special emphasis will be given on the numerical method used to perform the dynamics and the method's assumptions and limitations.  

Throughout the section, we will consider the following general class of lattice models,
\begin{align}
    \label{eqn:fullhamiltonian} \hat H & = \hat H_0(W) + U \hat H_{\mathrm{int}}, \\
    \hat H_0 & = \sum_{\langle \bm i, \bm j\rangle} A_{\bm i \bm j} \hat{a}_{\bm i}^\dag \hat{a}_{\bm j}, \\
    \hat H_{\mathrm{int}} & = \sum_{\langle \bm i, \bm j \rangle} D_{\bm i \bm j} \hat n_{\bm i} \hat n_{\bm j},
\end{align}
where $\hat a^\dag_{\bm i}$ ($\hat a_{\bm i}$) are the creation  (annihilation) bosonic/fermionic operators at site $\bm i$, $\hat n_{\bm i} = \hat a^\dag_{\bm i} \hat a_{\bm i}$. and $\langle \, \cdot\, , \,\cdot \,\rangle$ represent pairs of nearest-neighboring lattice sites. 
The non-interacting Hamiltonian $\hat H_0$ is determined by the single-particle Hamiltonian $A_{\bm i \bm j} = A_{\bm j \bm i}$, which has single particle eigenfunctions $\{\phi_l(\bm i) \}_l$ with energies $\{\lambda_l\}_l$. In the presence of sufficiently strong disorder these are Anderson localized, i.e.,  $\phi_l (\bm i) \sim e^{-|\bm i - \bm i_l|/\xi_{\text{loc}}}$, where ${\bm i}_l$ is the center of the $l$\textsuperscript{th} eigenstate. $\xi_{\text{loc}}(W) > 0$ denotes the localization length and is believed to be dependent on the disorder strength $W$ such that $\xi_{\text{loc}} \searrow 0$ as $W\rightarrow \infty$. Thus, $H_0$ can be diagonalized using Anderson creation (annihilation) operators, defined as $\hat \eta^\dag_l = \sum_{\bm i} \phi_l(\bm i) \hat c^\dag_{\bm i}$, such that $\hat H_0 = \sum_{\bm i} \lambda_l \hat I_l$, where $\hat I_l = \hat \eta^\dag_l \hat \eta_l$ is the non-interacting integral of motion. $\hat H_{\text{int}}$ is the interacting Hamiltonian, where $D_{\bm i \bm j} \sim \mathcal{O}(1)$. 

\subsection{$\ell$-bit phenomenology} \label{sec:lbitphenomenology}

In the limit of weak interactions and strong disorder ($U \ll W$), it is believed that the system is in the MBL phase and therefore can be described by an extensive number of local integrals of motion
\cite{basko2006metal, Aleiner2010, wahl2019signatures}, $\{\hat{\mathcal I}_l\}_l$, as 
\begin{equation} \hat H = \sum_l J^{(1)}_l \hat{\mathcal{I}}_l + \sum_{l,l'} J^{(2)}_{l,l'} \hat{\mathcal{I}}_l \hat{\mathcal{I}}_{l,l'} + \cdots. \end{equation}

In particular, for short-range interactions such as the case in Eq. \eqref{eqn:fullhamiltonian}, the interacting LIOMs are expected to be exponentially localized, that is, 
\begin{equation}
\frac{1}{\mathcal{N}} \text{tr}[\hat{\mathcal{I}}_l \hat A_{\bm i} ] \sim e^{-|{\bm i}_l-{\bm i}|/\upxi_{\text{loc}}},
\end{equation}
where $\hat A_{\bm i}$ is a generic observable with finite support real-space around site $\bm i$, $\mathcal{N}$ is the dimension of the Hilbert space, and $\upxi_{\text{loc}}(W, U) > 0$ is the interacting localization length. In the case of bosons, we can account for the unboundedness of the Hilbert space by replacing $\frac{1}{\mathcal N} \text{tr} [\; \cdot \;] \rightarrow \lim_{N\rightarrow \infty} \frac{1}{N^L} \text{tr}_{N} [ \; \cdot \;]$, where $\text{tr}_N$ denotes the trace over number of particles up to $N$. As a consequence, the couplings $J^{(2)}_{l,l'} \sim  \frac{1}{\mathcal{N}} \text{tr}[\hat{\mathcal{I}}_l \hat{\mathcal{I}}_{l'} \hat H ] \sim  e^{-|{\bm i}_l-{\bm i}_{l'}|/\upxi_{\text{loc}}} $ will also be exponentially localized.
Importantly, these exponentially decaying interactions imply a non-trivial propagation of information through the system \cite{nandkishore2015many}. 

It is also argued that the interacting eigenstates ($U \neq 0$) are adiabatically connected to the non-interacting ones \cite{Imbrie2016, Huse_phen_2014}. For example, in Ref. \onlinecite{Imbrie2016}, it has been analytically shown that at strong disorder and under some mild assumptions on the energy level-statistics, for a random spin chain in 1D, the system is diagonalized via a sequence of \textit{local} unitary operations. Crucially, the sequence of local unitary operations connect adiabatically the non-interacting eigenstates with the MBL ones, meaning that the exact eigenstates can be approximated at any target precision
using a  \textit{finite} sequence of local local unitary operations controlled by the perturbative parameter $\sim U/W$. 

\subsection{Constructing the effective $\ell$-bit Hamiltonian} \label{subsec:constructlbit}

Therefore, as a first approximation, in the limit of strong disorder and weak-interactions, we can approximate the interacting LIOMs $\{\hat{\mathcal{I}}_l\}_l$ with non-interacting ones $\{\hat I_l\}_l$, as follows. First, we write the Hamiltonian in terms of Anderson operators, i.e. 
\begin{align} \label{eqn:andersonbasishamiltonian}
\hat H & = \sum_l \lambda_l \hat \eta^\dag_l \hat \eta_l + \sum_{l,m,n,k} B_{lmnk} \hat \eta^\dag_l \hat \eta_m \hat \eta^\dag_n \hat \eta_k, \\
B_{lmnk} & = U \sum_{\langle \bm i,  \bm j \rangle }D_{\bm i \bm j} \phi_l(\bm i) \phi_m(\bm i) \phi_p(\bm j) \phi_q(\bm j).
\end{align}
We approximate the full Hamiltonian by only keeping terms of $\hat H_{\text{int}}$ that commute with the non-interacting integrals of motion, retrieving the effective $\ell$-bit Hamiltonian:
\begin{equation} \label{eqn:lbithamiltonian}
\hat H_{\ell\mathrm{-bit}} = \sum_l \epsilon_l \hat I_l + \sum_{l,m} B_{l,m} \hat I_l \hat I_m,\end{equation} where $\epsilon_l$ is the renormalized on-site energy, $B_{l,m} = B_{llmm} +s B_{lmlm}$ is the effective interaction term, and $s=1$ for bosons and $-1$ for fermions \cite{detomasi19, Wu_2019, Giu_alg_2019}. 

In Eq. \eqref{eqn:lbithamiltonian}, we discarded off-diagonal terms of $B_{lmnk}$. We are justified to do this as the $B_{lmnk}$ term involves the overlap of exponentially localized wavefunctions, and so any off-diagonal elements will be typically much smaller than the on-diagonal ones. We illustrate this below with an argument in 1D, although it is easily generalizable to 2D. Noting $\xi_\text{loc} \sim 1/\ln W$ for $W \gg 1$, we can expand $\phi_l(x)$ in powers of $W^{-1}$:
\begin{equation} \label{eqn:phitaylor} \phi_l(x) \sim \delta_{x_l, x} + W^{-1} (\delta_{x_l+1, x} + \delta_{x_l-1, x}) + \mathcal{O}\left(W^{-2}\right). \end{equation}
From Eq. \eqref{eqn:phitaylor}, we can see that if two indices are different in $B_{lmnk}$, we gain a factor of $W^{-1}$. Therefore the on-diagonal terms are of order $\sim U$, whilst the off-diagonal terms are either of order $U/W^2$ if two indices are same but rest are distinct or of order $U/W^4$ if all indices are distinct. 

The resulting dynamics of Eq. \eqref{eqn:lbithamiltonian} is that of first-order perturbation theory without considering corrections to the wavefunction. Interestingly, this can be seen as a version of Poincar\'e-Linstedt perturbation theory, in that only resonant corrections to the energy are considered. Unlike conventional perturbation theory, the Poincar\'e-Linstedt method generates approximate solutions are accurate for all times \cite{drazin1992nonlinear}. By disregarding the changes in the wavefunction, we avoid secular terms in the energies that would diverge over time, see Appendix \ref{apn:pl}.

\subsection{Efficiently calculating observables} \label{subsec:effcalcobs}

With the effective $\ell$-bit Hamiltonian $\hat H_{\ell\text{-bit}}$, any few-particle observable can be obtained from expectation values given by time evolution via a collection of effective quadratic Hamiltonians $\{ \hat F_{lm} \}_{l,m}$ which we will define below.
We note, however, that the underlying wavefunction is \textit{not} a Slater determinant, but a much more complicated object. 
We will first consider quadratic observables $\hat {\mathcal O} = \sum_{l,m} O_{l,m} \hat \eta^\dag_l \hat \eta_m$, whose expectation value can be obtained if all quadratic matrix elements $\langle \hat \eta^\dag_l \hat \eta_m \rangle_t$, or equivalently, $\langle \hat \eta_m \hat \eta^\dag_l \rangle_t$ are known, then generalize to higher order observables. For completeness, we provide the methods for both fermions and bosons.

\subsubsection{Product initial states}
First, consider an initial state that is a product state in bare-creation operators, $\lvert \psi_0 \rangle = \frac{1}{\sqrt{\prod_{\bm i} n_{\bm i}!}} \prod_{n=1}^N \hat a^\dag_{{\bm i}_n} \lvert 0 \rangle$.

It can be shown via elementary commutation relations that for the effective $\ell$-bit Hamiltonian Eq. \eqref{eqn:lbithamiltonian},
\begin{equation} \langle \hat \eta_m \hat \eta^\dag_l \rangle_t = e^{it(\Delta \epsilon_{lm} - B_{ll} + B_{mm})} \langle \hat \eta_m e^{it \hat F^{lm}} \hat \eta^\dag_l \rangle,
\end{equation}
where $\Delta \epsilon_{lm} = \epsilon_l - \epsilon_m$, ${\hat F^{lm} = \sum_{k} F^{lm}_k \hat I_k}$ is the effective quadratic Hamiltonian, whose matrix element is $F^{lm}_k = \tilde{B}_{lk} - \tilde{B}_{mk}$ and $\tilde B_{lm} = B_{lm} + B_{ml}$. Let us define $\mathcal{U}^\dag_t = e^{+it \hat F^{lm}}$. Using the property $\mathcal{U}^\dag_t \mathcal{U}_t = 1$ and inserting identities between the creation/annhilation operators, we find that
\begin{align} \label{eqn:prodstate}
    \langle \hat \eta_m e^{it \hat F^{lm}} \hat \eta^\dag_l \rangle = \langle 0 \rvert \hat a_{{\bm i}_N} \cdots \hat a_{{\bm i}_1} \hat \eta_m \hat \eta^\dag_{l,t} \hat a^\dag_{{\bm i}_1,t} \cdots \hat a^\dag_{{\bm i}_N,t} \lvert 0 \rangle,
\end{align}
where for any operator $\hat {\mathcal O}$, $\hat {\mathcal O}_t = e^{+it \hat F^{lm}} \hat {\mathcal O} e^{-it \hat F^{lm}}$. Because the time evolution is via a quadratic Hamiltonian, and the expectation is on the zero-particle state $\lvert 0 \rangle$, which is Gaussian for both fermions and bosons, Eq. \eqref{eqn:prodstate} can be evaluated using the following version of Wick's theorem, which we state for completeness. 

\textbf{Wick's theorem.} Let the bracket $\langle \; \cdot \; \rangle$ be an expectation value on a Gaussian state, and $\{ \hat A_i\}_{i=1,\dots,N}$, $\{\hat B^\dag_j\}_{j=1,\dots,N}$ be sets of annhilation and creation operators respectively, where each operator can be in different basis, and also be time-evolved via a quadratic Hamiltonian. Then expectation value of the product of these operators on the Gaussian state can be evaluated to be
\begin{equation} \left\langle \hat A_N \dots \hat A_1 \hat B^\dag_1 \dots \hat B^\dag_N \right\rangle = {\scriptstyle \begin{cases} \mathrm{perm} \\ \mathrm{det} \end{cases}} \! \! \! \! \! \! (\bm{M}) \quad \begin{cases} \mathrm{bosons} \\ \mathrm{fermions} \end{cases} \! \! \! \! \! \! \!,
\end{equation}
where the quadratic matrix elements are $[\bm M]_{ij} = \langle \hat A_i \hat B^\dag_j \rangle$. This similarly holds for the case $ \langle \hat A^\dag_N \dots \hat A^\dag_1 \hat B_1 \dots \hat B_N \rangle$, except with $[\bm M]_{ij} = \langle \hat A^\dag_i \hat B_j \rangle$.

This method of calculating observables for product states generalizes easily for higher-order expectation values, since we can always put them in a form where all creation operators are on the right, and the annhilation operators are on the left, after which we can apply Wick's theorem.
\subsubsection{Mixed initial states}
For fermions, expectation values for product initial states can be calculated via (Slater) determinants, which can be done in polynomial time. For bosons, however, this can only be be done using permanents, which scale exponentially with system size. We now show that if we instead consider mixed `thermal' Gaussian initial states, which are also more relevant in an experimental setting, where pure states cannot be prepared, we can compute the expectation values in polynomial time.

We consider the following initial density matrix,
\begin{equation} \label{eqn:thermalinit}
\hat \varrho_0 = {e^{- \hat Q}}/{\mathcal{Z}(\hat Q)},
\end{equation}
where $\hat Q = \sum_{\bm i} q_{\bm i} \hat n_{\bm i}$ some quadratic Hamiltonian and $\mathcal Z(\hat Q) = \text{tr} (e^{-\hat Q})$ is the partition function. In this case, expectation values can be calculated by a trace of the product of $\hat \varrho_0$ and the desired operator. First note that again using elementary commutation relations, $\langle \hat \eta^\dag_l \eta_m \rangle_t = e^{it(\Delta \epsilon_{lm} + B_{ll} - B_{mm} - \tilde{B}_{lm})}\langle e^{it \hat F^{lm} t} \hat \eta^\dag_l \hat \eta_m \rangle$. 

Thus, the expectation value for initial state Eq. \eqref{eqn:thermalinit} can be found to be
\cite{levitov1996pis, klich2002full}
\begin{equation} \label{eqn:thermalquadratic} \langle e^{-it \hat F^{lm} t} \hat \eta^\dag_l \hat \eta_m \rangle = \frac{
    \left[(\bm{1}+s {\bm{f}}' - s { \bm{f}}' e^{it\bm{F}^{lm}})^{-1}
    {\bm{f}}' e^{it \bm{F}^{lm}}
    \right]_{ml}
}{
    \det \left( \bm{1} + s {\bm{f}}' - s {\bm{f}}' e^{it \bm{F}^{lm}}\right)^s
},\end{equation}
where $s=+1$ for bosons, and $-1$ for fermions, $\bm F^{lm} = \mathrm{diag} (\{ F^{lm}_k \}_k)$, $\bm{f} = \mathrm{diag}\left(\left\{\frac{e^{- q_{\bm i}}}{1 - s e^{- q_{\bm i}}}\right\}_{\bm i}\right)$, and the dash operator $'$ is defined to rotate a matrix into the $\ell$-bit basis, so that for any matrix $\bm M$,  $[\bm{M}']_{lm} = \sum_{\bm i, \bm j} \phi_l(\bm i) \phi_l(\bm j) [\bm M]_{\bm i \bm j}$.

Eq. \eqref{eqn:thermalquadratic} can be derived as follows. First, note that 
\begin{equation}\langle e^{it \hat F^{lm}} \hat \eta^\dag_l \hat \eta_m \rangle = \text{tr} \left( \frac{e^{-\hat Q}}{\mathcal{Z}(\hat Q)} e^{it \hat F^{lm}} \hat \eta^\dag_l \hat \eta_m \right).\end{equation}
Because both $\hat Q$ and $\hat F^{lm}$ are quadratic operators, via the Baker-Campbell-Hausdorff formula \cite{rossmann2006lie}, there exists some quadratic operator $\hat C = \sum_{k} C_k \hat {\#}_k$, where $\hat {\#}_k = \hat \gamma^\dag_k \hat \gamma_k$, and $\hat{\eta}_l = \sum_{a} [\bm{V}]_{l a} \hat{\gamma}_a$ for some matrix $\bm V$.
Therefore 
\begin{equation}\text{tr} \left( e^{-\hat Q} e^{it \hat F^{lm}} \hat \eta^\dag_l \hat \eta_m \right)
= 
\sum_{ab} [\bm{V}]^{\dagger}_{i a} [\bm{V}]_{m b} \;
\text{tr}\left( e^{\sum_k C_k \hat{\#}_k} \hat{\gamma}^{\dagger}_a \hat{\gamma}_b \right).\end{equation}
Since the trace vanishes whenever $a \neq b$, the trace is a standard bosonic or fermionic counting problem, evaluated to be
\begin{equation}\begin{aligned} \text{tr} \Big( e^{\sum_k C_k \hat{\gamma}_k^\dagger \hat{\gamma}_k} & \hat{\gamma}^{\dagger}_a \hat{\gamma}_b \Big) 
= \delta_{ab} \sum_{\{\vec{\#}\}} e^{\sum_k C_k \#_k} \#_a \\ 
= \delta_{ab}
\det & \left( \bm{1} - s e^{\bm{C}} \right)^{-s} \left[ (\bm{1}-s e^{\bm{C}})^{-1}e^{\bm{C}} \right]_{aa},\end{aligned}\end{equation}
where $\bm C = \mathrm{diag}(\{C_k\}_k)$. Substituting and using the summation over $\bm V$ matrices to transform back into the original basis, we find that
\begin{equation}
\begin{aligned}
    \text{tr} \Big( e^{- \hat{Q}} & e^{+it\hat{F}^{lm}} \hat{\eta}^{\dagger}_l \hat{\eta}_m \Big) \\ & = \text{det}\left( \bm{1} - se^{-\bm{Q}'} e^{it \bm{F}^{lm}}\right)^{-s} \\ & \times \left[ (\bm{1}-se^{-\bm{Q}'} e^{it \bm{F}^{lm}})^{-1} e^{- \bm{Q}'} e^{it \bm{F}^{lm}} \right]_{ml},
\end{aligned}
\end{equation}
where $\bm Q = \mathrm{diag}(\{ q_{\bm i} \}_{\bm i})$. Similarly, The partition function can be  evaluated to be $\mathcal{Z}(\hat Q) = \det \left\{\bm{1} - s e^{- \bm{Q}} \right\}^{-s}$. Finally, we can substitute the definition of $\bm f$ to find Eq. \eqref{eqn:thermalquadratic}.

Since only a determinant is required, we note that for the initial state in Eq. \eqref{eqn:thermalinit}, the computational power required scales polynomially with $L$ and  is independent of the targeted time. Similarly, for higher-order expectation values, we can simply represent the expectation in a form such that all the creation operators to the left, and the annhilation operators to the right. Since the expectation is over a Gaussian state time-evolved by an effective quadratic Hamiltonian, we can use Wick's theorem to simplify into quadratic expectation values, then use Eq. \ref{eqn:thermalquadratic} to calculate the result in polynomial time.

By substituting $t=0$ into Eq. \eqref{eqn:thermalquadratic}, we can see that we can control the initial expectation value of occupation, $f_{\bm i}$, by appropriately choosing the quadratic Hamiltonian $\hat Q$ as
\begin{equation}
    f_{\bm i} = \frac{e^{-q_{\bm i}}}{1 - s e^{- q_{\bm i}}}.
\end{equation}
The initial density matrix is then $\hat \varrho_0 = \sum_{\{ \bm n\}} w(\bm n) \left \lvert \bm n \right \rangle \left \langle \bm n \right \rvert$, where $\{ \bm n\}$ is the set of all possible real-space occupations, and the weights of each occupation is $w(\bm n) = \prod_{\bm i} w(n_{\bm i})$, which can be separated into weight of each site as $w(n_{\bm i}) = f_{\bm i}^{n_{\bm i}} / (1 + s f_{\bm i})^{n_{\bm i} + s}$, which in the case of bosons, is exponentially decaying with occupation, with the rate controlled by $f_{\bm i}$.

\subsection{Range of validity in the case of bosons} \label{subsec:rangeofvalidity}

Having introduced our method, let us now take the chance to explore its range of validity.

With this method we can compute observables for bosons in regimes that are far beyond the capabilities of other computational methods such as exact diagonalization or tensor network methods in terms of time $t$, system size $L$ or number of particles $N$. Consequently, we cannot directly compare with such an exact reference. Still, our method is expected to be accurate and controlled as we now aim to argue.

In the seminal work of Basko, Aleiner and Altshuler \cite{BASKO20061126}, and confirmed by others \cite{Ros2015, Imbrie2016} later, it  has been shown that perturbation theory starting from the Anderson eigenstates is stable as long as interactions are sufficiently weak. This stability is deeply rooted in the definition of MBL, which implies that integrals of motions are adiabatically connected to the non-interacting ones. In particular, this stability is a consequence of the existence
of the spatioenergetic anti-correlations between eigenstates in the single-particle Anderson model. Single-particle eigenstates that are close together spatially are in general very different in terms of energy. As a consequence, even if denominators of energy differences are small in perturbation theory, the full perturbative corrections are generally nevertheless suppressed exponentially due to the exponentially small overlap of the tails of the involved single-particle wave functions. 

This adiabatic continuity to the noninteracting limit forms the basis for our presented method. On the one hand this allows us to choose as the integrals of motion the noninteracting ones to leading order. On the other hand, the off-diagonal interaction terms can, again to leading order, be discarded as they introduce only weak perturbative corrections, see also the case of fermions~\cite{detomasi19}. Overall, this makes our approach controlled in the limit of strong disorder provided interactions are sufficiently weak. Let us just note, however, that bosons are always expected to become ergodic at sufficiently high energies~\cite{Aleiner2010}, which naturally limits the occupation numbers, i.e., the density of bosons, in our considered systems. 

Concerning the dynamics addressed in this work it is, however, also a natural question up to which time scale the solution can be considered accurate. For fermions it has been shown already that the approach yields controlled results for all times by comparing to an exact reference~\onlinecite{detomasi19}. For bosons we don't have such an exact reference at hand. However, it is nevertheless possible to estimate the temporal validity. In the end our approach becomes uncontrolled whenever resonant processes, beyond the ones already taken care of, contribute significantly. From the simple perturbative analysis in Sec. II.B we have seen already that such resonances don't appear in low-order perturbation theory. Of course, this a priori doesn't exclude significant resonances at higher orders, although it has been argued in a previous work that an MBL phase is supposed to exist at sufficiently low energy densities~\cite{Aleiner2010}. Still, even if resonances occur at higher orders in perturbation theory these would only contribute at longer times so that up to that point our description would still be accurate and controlled.


Let us finally emphasize again that in this work we are aiming at the localization dynamics of a 2D bosonic system at strong disorder. It is important to point out that the existence of a \textit{genuine} MBL phase in 2D is not well established as compared to the 1D case. On the one hand, recent experiments and numerical simulations have shown the stability of an MBL phase in 2D systems \cite{Choi1547, kennes2018manybody,Th_veniaut_2020,Altman_2019,chertkov2020numerical,Francesca_2021,doggen2020slow}. On the other hand, general considerations which take into account rare non-perturbative effects have laid out the possibility that an MBL phase in dimension higher than one might not exist asymptotically at large length and time scales \cite{Thiery_2018, Gapa_2019,Insta_2017,De_Roeck_2017}. Importantly, these non-perturbative effects evoke the existence of entropic ergodic bubbles, which destabilise the MBL phase at extremely long times \cite{Thiery_2018, Gapa_2019,Insta_2017,De_Roeck_2017, DeTo2020, Kos_2019}. As a result, even if MBL in 2D might finally turn out to be unstable, we expect that our approach based on the existence of the LIOMs at strong disorder and weak interactions will still be accurate and controlled for fairly long time scales before the instability might kick in.

\section{Bosonic Many-Body Localization} \label{sec:bosonicmbl}

Having elucidated the main theory of our method, we now apply it to a specific case, namely to a disordered Bose-Hubbard model in 1D and 2D,
\begin{equation}
    \hat H = \sum_{\langle \bm i, \bm j \rangle} \left( \hat a^\dag_{\bm i} \hat a_{\bm j} + \mathrm{h.c.} \right) + \sum_{\bm i} h_i \hat n_{\bm i} + U \sum_{\bm i} \hat n_{\bm i} \left( \hat n_{\bm i} - 1 \right),
    \label{eqn:bosehubbard}
\end{equation}
where $\hat a^\dag_{\bm i}$ ($\hat a_{\bm i}$) are the creation  (annhilation) bosonic operators at site $\bm i$. $\{h_{\bm i}\}_{\bm i}$ is a random field between $[-W, W]$ and $U$ is the interaction strength.  

We consider an initial mixed state, which is close to a charge-density wave state in the sense of $f_{(x=2k,y)}=1$, $f_{(x=2k+1,y)}=0$ in Eq. \eqref{eqn:thermalquadratic}. We study the system both in a 1D lattice of size $L$ and a rectangular lattice (2D) of size $S = L \times L/2$. Both cases are subject to open boundary conditions. Moreover, we focus our study in the strong disorder and weak interactions ($U=0.1$) regime. Here, we expect the aforementioned method to provide an accurate description over several orders of magnitude in time \cite{detomasi19}.  

\subsection{Temporal fluctuations} \label{subsec:temporalflucs}

One of the crucial difference between an MBL phase and an Anderson localized one is the fact the system relaxes in the long-time limit, even though it does not thermalize. This equilibration is manifest in the expectation values of local observables, which relax to stationary values at long times in the thermodynamic limit, and show decaying temporal fluctuations. To explore this difference, in this subsection, we study the temporal fluctuations of the disordered Bose-Hubbard Hamiltonian.

For concreteness, We quantify the temporal fluctuation in the system with the following observable,
\begin{align}
\Delta n^2 (t) = \frac{1}{L} \sum_x \Delta n^2_x(t), \; \; \Delta n_x^2(t) = \left( \left\langle \hat n_x(t) \right\rangle - \left \langle \hat n_x \right\rangle_{\mathrm{tav}} \right)^2,
\end{align}
where $\langle \; \cdot \;\rangle_{\text{tav}}$ indicates the long-time average, $x$ denotes the spatial site index in the x-direction in both 1D and 2D, and $y=0$ for 2D. 

It has been argued \cite{detomasi19, Quantum_serby_2014}, that the temporal fluctuations decay algebraically in time in an MBL phase in 1D, $\Delta n^2 (t)\sim t^{-\alpha}$. In particular, it has been predicted that the exponent $\alpha$ is proportional to the localization length ($\alpha \propto \xi_{\text{loc}}$). 

Here we apply our method to numerically study the decay of time-fluctuations of local observable in 1D and 2D. We confirm the algebraic decaying in 1D and we refine this result in 2D by showing that it may relax as 
\begin{equation}
    \Delta n^2(t) \sim e^{- \beta \ln^2(t)}.
\end{equation}
In particular, we present an analytical argument using the effective model and propose a general asymptotic form for the decay of the temporal fluctuation in any dimensions $d$, $\Delta n^2(t) \sim e^{-\beta_d \ln^d(t)}$. This result, as well as the further analytic predictions that we will present, shows that our method allows us to not only numerically compute observables at regimes inaccesible to other methods, but also allows us to analytically explain the behaviour of these observables.

\begin{figure}[ht] 
\includegraphics[width=1.0\columnwidth]{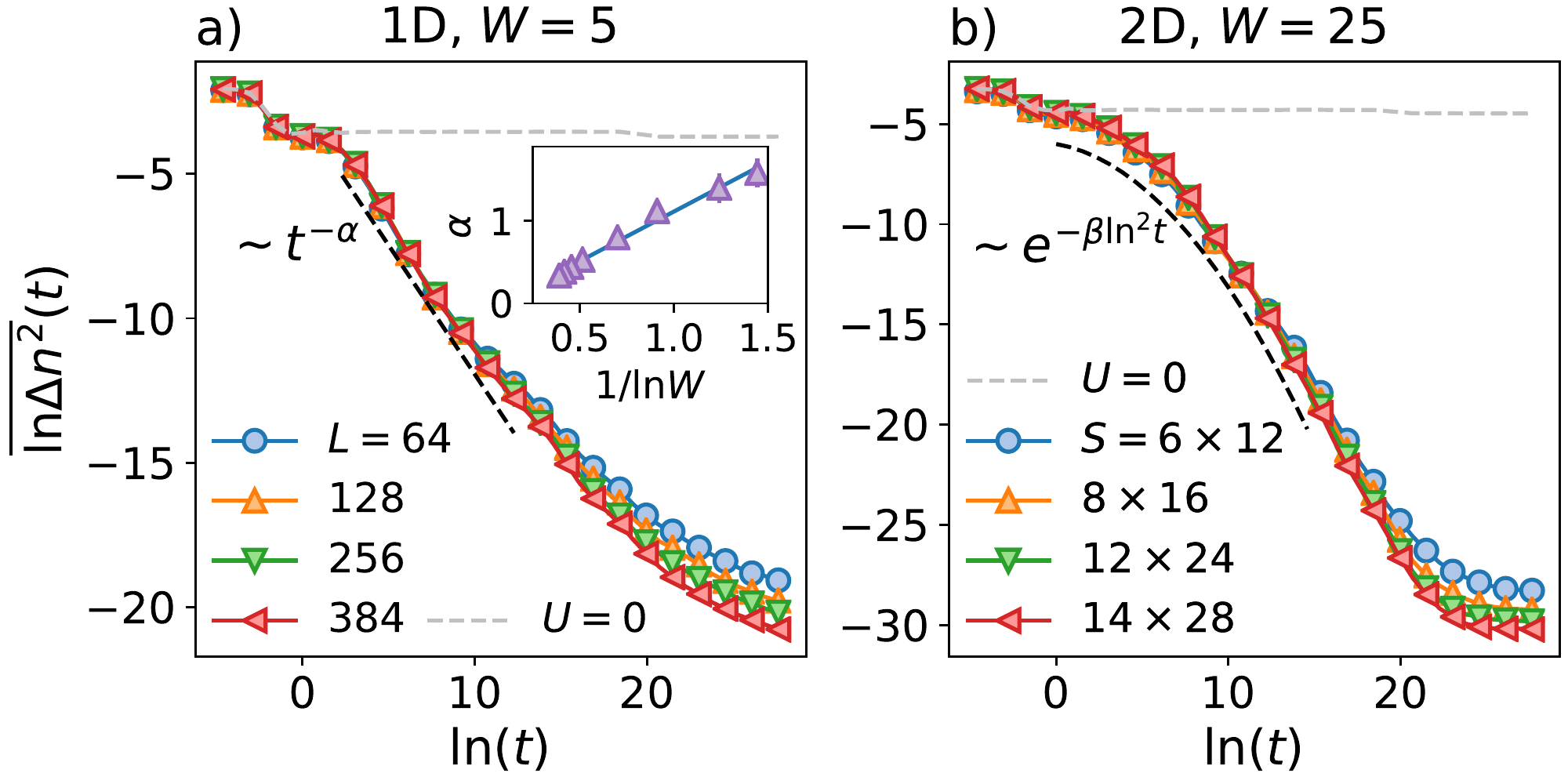}
\caption{\label{fig:timeflucs} Disorder-averaged temporal fluctuations $\overline{\ln \Delta n^2}(t)$ for (a): 1D and (b): 2D. Non-interacting cases are shown as a grey dashed line. In the presence of interactions, the plot is consistent with a power-law decay in the case of 1D, and a quasi-polynomial decay in the case of 2D, as indicated by the fit using black dashed lines. Inset of (a): linear fit on scaling of the power law $\alpha$ with $1/\ln W  \sim \xi_{\mathrm{loc}}$ for the 1D system.}
\end{figure}

Figure \ref{fig:timeflucs} (a), (b) shows $\Delta n^2(t)$ for the 1D and the 2D case, respectively, where $\overline{\; \cdot \;}$ denotes disorder averaging. We can see that in both 1D and 2D, in the interacting case, $\Delta n^2(t)$ decays with time, unlike the non-interacting case $U=0$ (dashed lines). The decay for the 1D case is consistent with an algebraic decay $\Delta n^2 (t)\sim t^{-\alpha}$, that is the dynamical range of time for which a polynomial decay is visible increases with system size (see Fig. \ref{fig:timeflucs} (a)). In the inset of Fig. \ref{fig:timeflucs} (a), we can also see that $\alpha \propto 1/\ln W$, providing evidence that $\alpha \propto \xi_\text{loc}$. In the 2D case, we observe deviations from a pure algebraic decay, and see that the relaxation is consistent with  $\Delta n^2(t) \sim e^{-\beta \ln^2(t)}$. 
 
We now present an analytical argument to support the found decay for $\Delta n^2(t)$. First, we assume that the qualitative quench behaviour will not change given similar initial conditions. So, we choose the initial state that corresponds to taking the trace over the $(0,1)$ site occupation sector, 
\begin{equation} \label{eqn:01trace}
\lvert \psi_0 \rangle = \left(\prod_l \frac{1 + \hat \eta^\dag_l}{\sqrt{2}} \right) \lvert 0 \rangle = 2^{-S/2} \sum_{ \left\{ I_i | I_i \in \{0,1\} \right\} } \lvert \vec I \; \rangle.
\end{equation}
On this state, we find that 
\begin{equation} \label{eqn:timeflucapprox} \Delta n_x^2(t) \sim \left(\sum_{l\neq m} \phi_l(x) \phi_m(x) e^{it \Delta \tilde{\epsilon}_{lm}} \prod_{k \neq l, m} \cos (F^{lm}_k t/2) \right)^2, \end{equation}
where $\tilde{\epsilon}_q = \epsilon_q +B_{qq} + \sum_{k\neq l,m} \tilde{B}_{qk}/2$. Since $F^{lm}_k \sim U e^{- \text{min}(\lvert {\bm i}_l-{\bm i}_k \rvert,\lvert {\bm i}_m-{\bm i}_k \rvert)/\xi_{\text{loc}}}$, factors in the cosine product such that ${\bm i}_k$ is further away from ${\bm i}_l$ or ${\bm i}_m$ than $\xi_\text{loc}$ will be nearly $1$, we can restrict the limits product to over $k$ that meets the condition $1 \lesssim U e^{- \text{min}(\lvert {\bm i}_l-{\bm i}_k \rvert,\lvert {\bm i}_m-{\bm i}_k \rvert)/\xi_{\text{loc}}} t / 2 $, and approximate the factors as some constant less than one. The number of such factors is approximately the volume of two $d$-balls of radius $\xi_{\text{loc}} \ln (Ut)$. Therefore, there are $\sim \xi_{\text{loc}}^d \ln^d (Ut)$ such factors, which points towards the quasi-polynomial decay, as claimed \cite{Shreya_2017, detomasi19}. 


\subsection{Two-point correlators} \label{subsec:fluctwotime}

In this subsection, we study the behavior of two-point correlator function, the fluctuation of the two-time commutator. The two-time commutator of local observables, such as $\left[\hat n_x(t), \hat n_0 \right]$, is a key object that describes spatiotemporal propagation of quantum correlations, and can be used to study the spatial growth of local operator such as $\hat n_x(t)$, initially localized at site $x$. This object has also been studied extensively in the context of Lieb-Robinson bounds \cite{lieb72}. To get a feeling for the typical behaviour of its expectation value, we again evaluate on the initial state Eq. \eqref{eqn:01trace}, and find that $\langle \left[\hat n_x(t), \hat n_0 \right] \rangle \sim \sum_{l,m,n,k} M_{lmnk}$, where the typical matrix element looks like 
\begin{equation}\begin{aligned}M_{lmnk} \sim \phi_l(x) \phi_m(x) \phi_n(0) \phi_k(0) \left(e^{it F^{lm}_n} - e^{it F^{lm}_k}\right) \\ \times e^{it\Delta \tilde{\epsilon}_{lm}} \prod_{q\neq l,m,n,k} \cos(F^{lm}_q t/2),\end{aligned}\end{equation}
where $\tilde{\epsilon}_q = \epsilon_q +B_{qq} + \sum_{k\neq l,m,n,k} \tilde{B}_{qk}/2$. The difference in complex exponential factor suggests that after disorder averaging, the quantity will average to zero. Hence, we can study its fluctuation, a two-point correlator ${\mathcal{G}}(x,t)$:
\begin{align}
{\mathcal{G}}(x,t) = - \left\langle \left[\hat n_x(t), \hat n_0 \right] \right\rangle^2.
\end{align}
From cosine factor of the typical matrix element, we can use a similar argument as that of $\Delta n^2(t)$ to argue that ${\mathcal{G}}(x,t)$ too will show a quasi-polynomial decay at moderate times. 

\begin{figure}[ht]
\includegraphics[width=1.0\columnwidth]{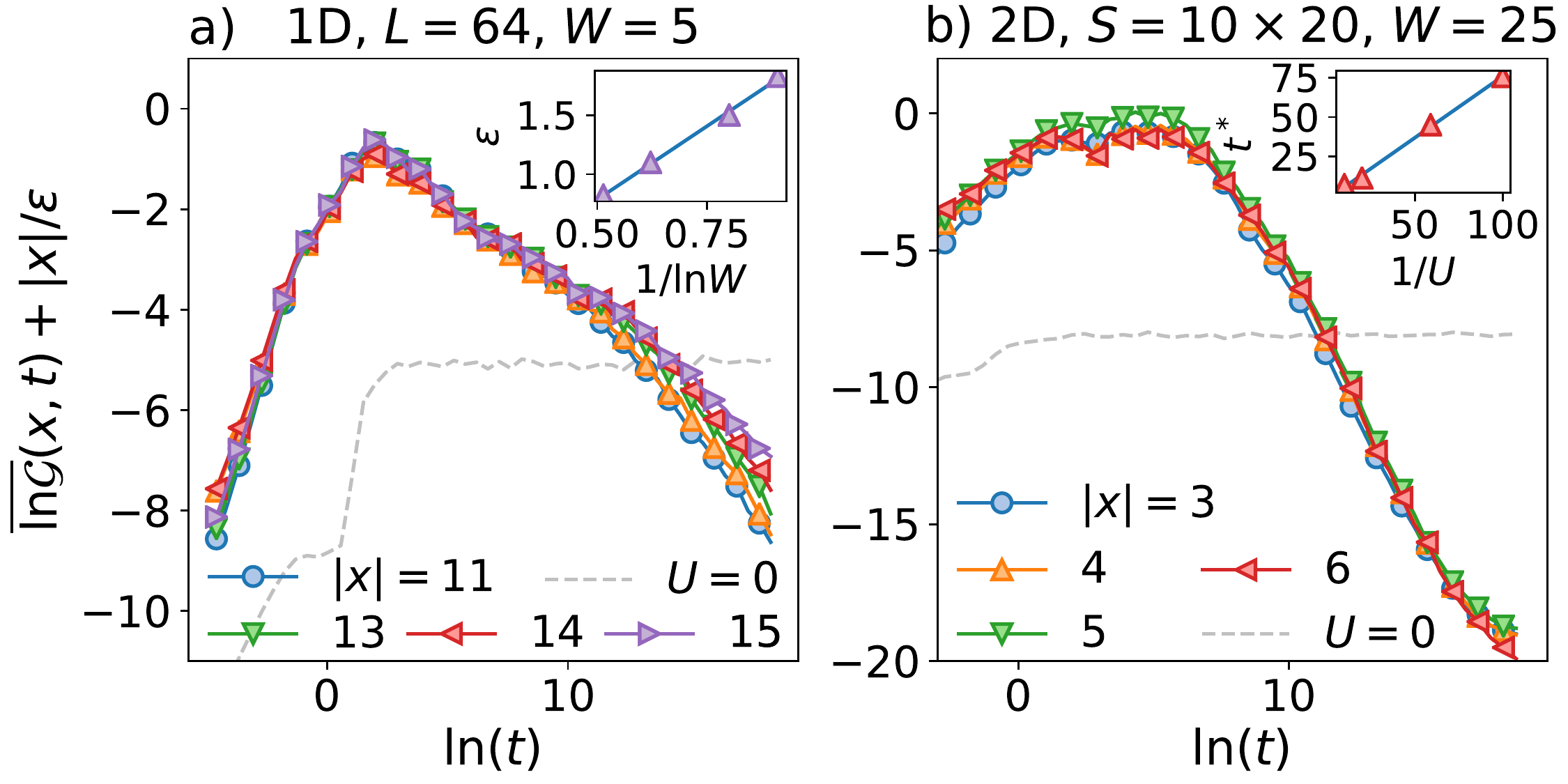}
\caption{\label{fig:Glines} Collapsed plot of the disorder-averaged two-time correlator $\overline{\ln \mathcal{G}}(x,t)$ for (a): 1D, and (b): 2D, by addition of $\lvert x \vert / \epsilon$, where $\epsilon=0.9$ for 1D, and $3.5$ for 2D. Inset of (a): linear fit on scaling of decay length scale $\epsilon$ with $1/\ln W  \sim \xi_{\mathrm{loc}}$ for the 1D system. Inset of (b): linear fit of peak time $t^*$ with $1/U$ for the 1D system.}
\end{figure}

In the 1D case, Fig. \ref{fig:colorplots} (c) shows a color plot of ${\mathcal{G}}(x,t)$,
from which we can deduce its qualitative properties. In the 1D case, Fig. \ref{fig:colorplots} shows that in the case of the AL phase, we see an initial spread of fluctuations up to localisation length, followed by a static phase, which persists over many decades. Meanwhile, the MBL phase has a different qualitative behaviour; at small times, $\mathcal{G}(x,t)$ spreads and grows, but unexpectedly, starts to decay at moderate times \cite{detomasi19}.
The quantitative growth of ${\mathcal{G}}(x,t)$ is shown in Fig. \ref{fig:Glines} (a). In the AL case, we see that there is an initial growth phase followed by saturation. For the MBL case, however, there exists a time $t^*$, around which the initial growth phase turns into what appears to be a polynomial decay, as predicted by the analytic argument. We expect that $t^* \sim 1/U$, on dimensional grounds and the fact that the peak is due to interactions, since it does not exist in the Anderson-localized case. This argument is supported by the inset of Fig. \ref{fig:Glines} (a). Additionally, the data collapse from adding a factor of $\lvert x\rvert$ from  ${\mathcal{G}}(x,t)$ indicates that the magnitude of ${\mathcal{G}}(x,t)$ decays exponentially in space, i.e. ${\mathcal{G}}(x,t) \sim e^{- \lvert x \rvert/ \epsilon} e^{-C_d \xi_\text{loc}^d \ln^d (Ut)}$. We expect that $\epsilon \sim \xi_\text{loc}$, as $\xi_\text{loc}$ is the only relevant length scale of the system. This is again supported by the inset of Fig. \ref{fig:Glines} (b). Finally in Fig. \ref{fig:colorplots} (d) and Fig. \ref{fig:Glines} (b) show the same phenomena, but for the 2D case, except for the case of Fig. \ref{fig:Glines} (b), where the decay is consistent with a quasi-polynomial decay, i.e. $\sim \ln^2 (t)$ for $\ln (t) \gtrsim 0$.

\subsection{Out-of-time-order correlators} \label{subsec:otocs}

\begin{figure}[ht]
\includegraphics[width=1.0\columnwidth]{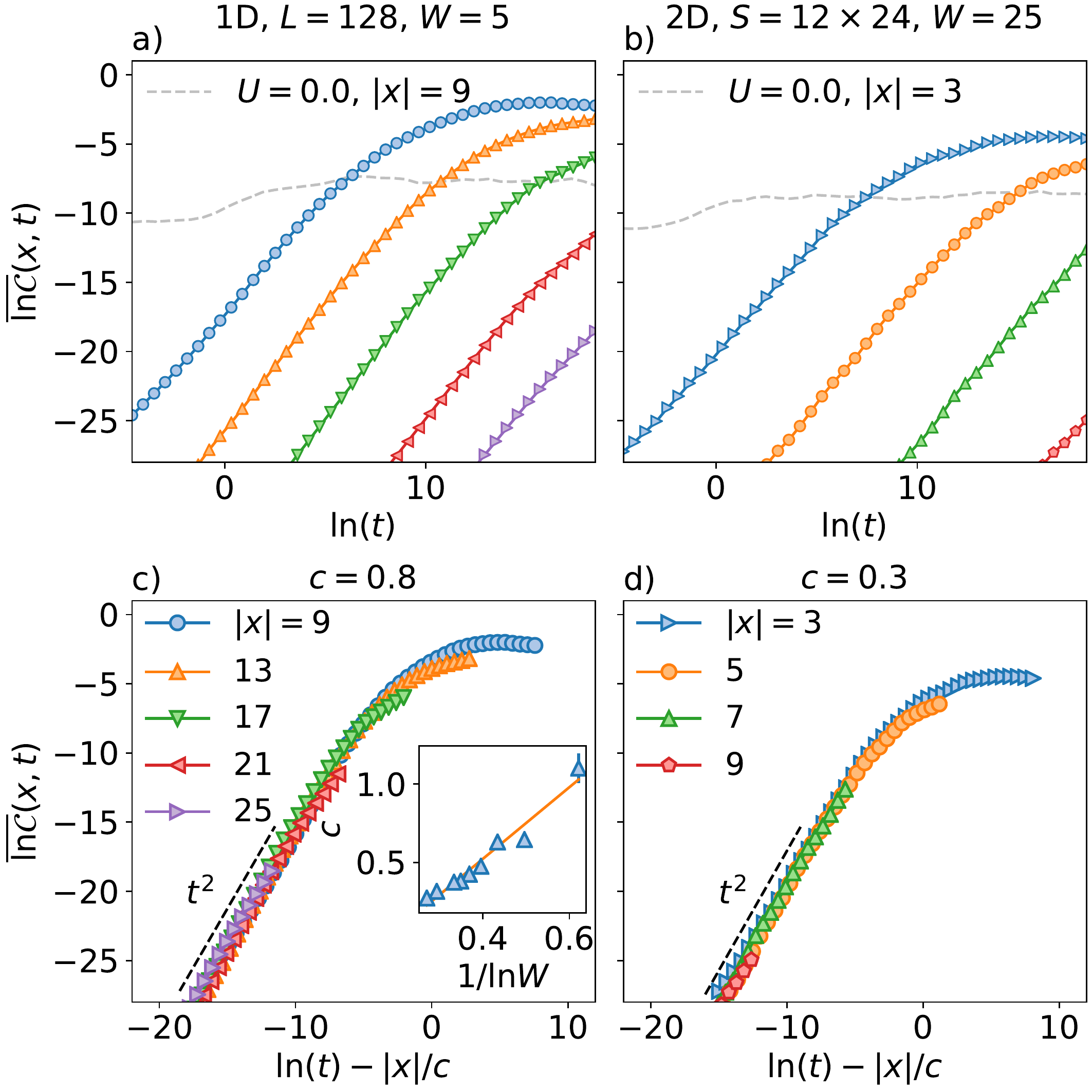}
\caption{\label{fig:otoclines} (a), (b): Disorder averaged OTOC $\overline{\ln \mathcal{C}}(x,t)$ for 1D and 2D, respectively.  (c), (d): Collapse of OTOCs by using the logarithmic light cone. MBL case shows $\sim t^2$ power law growth, as indicated by a fit in black dashed lines. This is absent in the AL case (grey dashed-lines). Inset of (c): linear fit of light cone log-velocity $c$ with $1/\ln W \sim \xi_{\mathrm{loc}}$.}
\end{figure}

In this subsection, we inspect the behavior of the OTOC in 1D and 2D. The OTOC has been widely studied in the context of MBL \cite{lee19, chen2017out} and is thought to describe quantum information scrambling in many-body systems, and is known to diagnose the onset of chaos the semi-classical limit \cite{larkin1969quasiclassical, maldacena2016bound, huang2017out}.
It is also the second moment of the two-time commutator discussed in the previous subsection and is defined as
\begin{equation} \label{eqn:otoc}
    {\mathcal{C}}(x,t) = \left\langle \left[ \hat n_x(t), \hat n_0 \right]^\dag \left[ \hat n_x(t), \hat n_0 \right] \right\rangle.
\end{equation}
The OTOC, which is 8th-order in creation and annhilation operators, has been numerically simulated for spin chains and fermions, but never for a bosonic system at the time of writing to the best of our knowledge.
Our simulation of the OTOC shows that our method can be used to simulate  high-order correlation functions. Fig. \ref{fig:colorplots} (a), (b) shows a color plot of ${\mathcal{C}}(x,t)$ for 1D and 2D respectively. We see that for the AL case, the OTOC spreads only up to the localisation length then becomes stationary - correlations no longer spread over time. On the other hand, the MBL phase shows stark difference to the AL phase, where we can clearly see a logarithmic light-cone that persists even at very large times, indicating that quantum correlations continue to spread at constant `log-velocity' $c$ with $\ln (t)$. 
In Fig. \ref{fig:otoclines} (a), we see that the OTOC first has a growth phase at a power law of $\sim t^2$, then then saturates to a constant value at long times. This agrees with our analytic prediction, which we will present below, as well as those of spin chains \cite{lee19, swingle2017}, providing evidence that OTOCs in bosonic systems share similar features to those of spin chains. The data collapse by shifting of time via log-velocity $c$, as shown in Fig. \ref{fig:otoclines} (c), confirms that the spread is indeed linear in $\ln(t)$. In the inset of Fig. \ref{fig:otoclines} (c), we study how the log-velocity $c$ scales as a function of disorder strength $W$. We see that the log-velocity scales as $1/\ln W $, which, since $\xi_{\text{loc}} \sim 1/\ln W$ for strong disorder, means that increasing localization length increases the log-velocity of the light cone. Much of the same observations can be seen in 2D from Figs. \ref{fig:colorplots} (b), \ref{fig:otoclines} (b) and \ref{fig:otoclines} (d).
We now provide arguments for the observed spatiotemporal behaviour of the OTOCs. In a non-interacting localized phase, we can expect that after an initial spread, ${\mathcal{C}}(x,t)$ will have support for $\lvert x \rvert \lesssim \xi_\text{loc}$. This is because the non-interacting operators $\hat \eta^\dag_l = \sum_{\bm i} \phi_l (\bm i) \hat c^\dag_{\bm i}$, which the system evolves with, only have support on sites $\xi_\text{loc}$ around ${\bm i}_l$, which is what we saw. On the other hand, in the interacting phase, we expect that $\hat n_x(t)$ will have significant support at site $0$ only when $t \sim 1/B_{x0}$. Since $B_{lm} \sim  U e^{-\lvert {\bm i}_l - {\bm i}_m \rvert / \xi_\text{loc}}$, it suggests non-zero ${\mathcal{C}}(x,t)$ for $\lvert x\rvert \lesssim \xi_\text{loc} \ln (Ut)$. Thus, we expect a logarithmic light-cone $\sim \ln (t)$, with the log-velocity scaling with localization length $c \sim \xi_\text{loc}$, as observed. We suspect that the reason this argument does not hold for ${\mathcal{G}}(x,t)$ is because the decay process is the faster, more dominant process, compared to the slow spread of correlation. We can also argue the early-time growth of the OTOC should go as $\sim t^2$, as follows. First, we assume that the qualitative behaviour should not depend on the choice of the local operators, as long as they are not correlated at $t=0$. So, we choose the commutator $[ \hat{I'}_{l}(t), \hat I_0]$, where $\hat{I'}_{l} = \sum_{m,k \in \{l, l+1\}} \hat \eta^\dag_m \hat \eta_k$, such that $\lvert {\bm i}_l \rvert \geq \xi_\text{loc}$. Next, we argue that the early time scaling behaviour can be deduced from one of the terms, since the OTOC is just a sum of them. Choosing one of the terms, it can be shown that 
\begin{equation}\langle \hat{I'}_{l}(t) \hat I_0 \hat{I'}_{l}(t) \hat I_0 \rangle \sim \cos(\Delta \tilde{\epsilon}_{l,l+1} t) \prod_{k \neq 0, l, l+1} \cos (F^{l,l+1}_k t /2),\end{equation} 
where $\tilde{\epsilon}_q = \epsilon_q + B_{qq} + \sum_{k \neq 0, l, l+1} \tilde{B}_{kq} /2 + \tilde{B}_{q0}$. Upon expanding this expression for small $t$, we retrieve the $\sim t^2$ growth at early times, consistent with the observed result.

\subsection{Dependency on occupation amplitudes} \label{subsec:occamp}

\begin{figure}[ht]
\includegraphics[width=1.0\columnwidth]{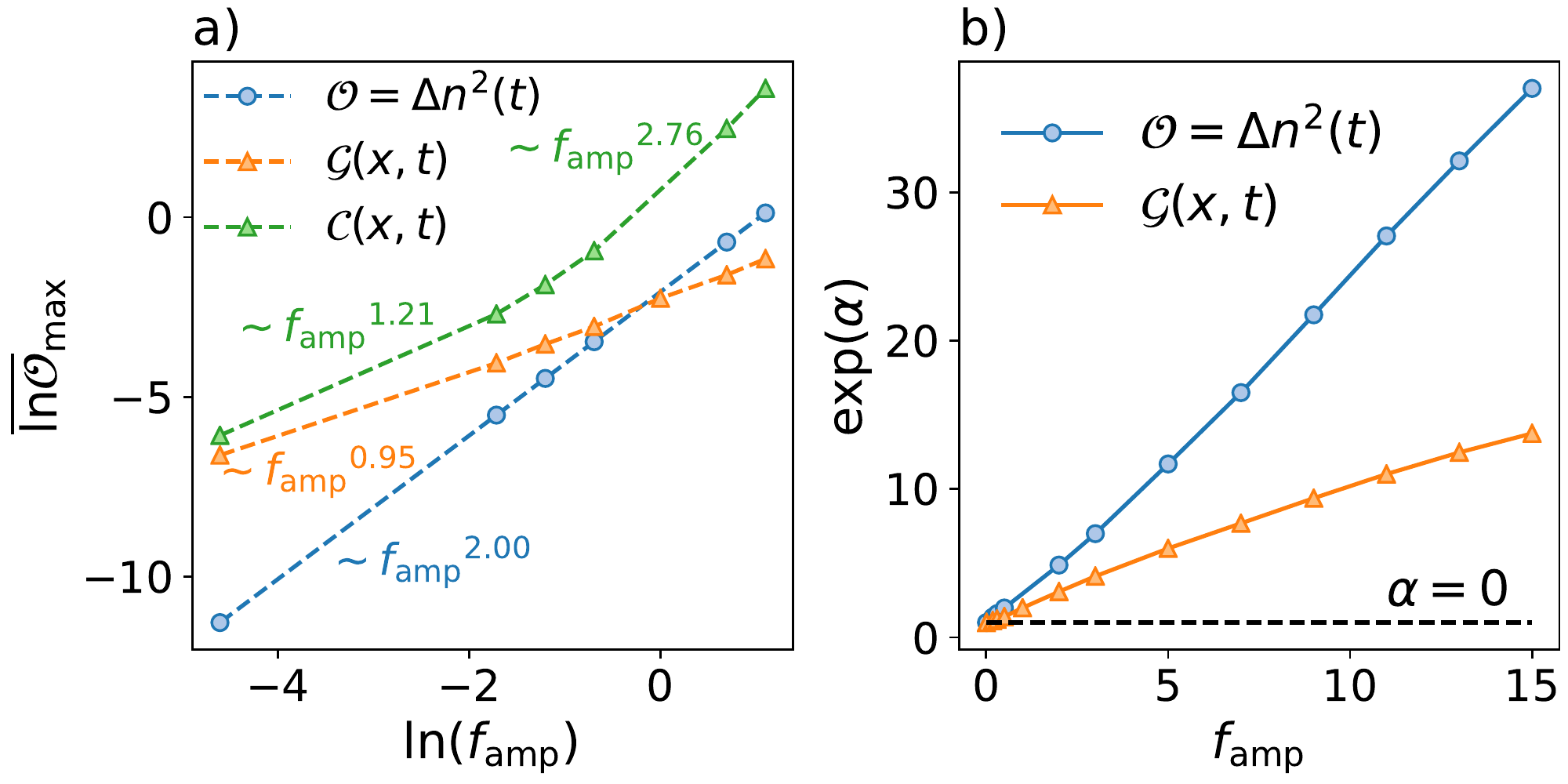}
\caption{\label{fig:amps_slopes} (a): Scaling of the disorder-averaged magnitudes of observables $\overline{\ln \mathcal{O}}_{\max}$ with the occupation amplitude $f_\text{amp}$. (b): Scaling of the power-law decay $\alpha$ with $f_\text{amp}$ for the 1D case. The black dashed line indicates the point $\alpha=0$.}
\end{figure}

Now that we have studied properties shared between bosonic, fermionic and spin-chain systems, we now consider the effect of the occupation amplitude $f_\text{amp}$ on the studied observables $\{\mathcal O\}$, a property unique to bosons. We consider N\'eel-like initial states with various amplitudes $f_{(x=2k,y)}=f_\text{amp}$, $f_{(x=2k+1,y)}=0$. First, we study how the maximum amplitude the observables, $\mathcal{O}_{\max},$ changes with $f_\text{amp}$. Consider the observable $\Delta n^2(t)$. For the case $f_\text{amp} \ll 1$, we can expand Eq. \eqref{eqn:thermalquadratic} by using the properties $\det (\bm 1 + \delta \bm A) = 1 + \delta \text{tr} \bm A + \mathcal{O}(\delta^2)$ and $(\bm 1 + \delta \bm A)^n = \bm 1 + n \delta \bm A + \mathcal{O}(\delta^2)$, to find that quadratic observables should scale as $\sim f_\text{amp}$. For the case $f_\text{amp} \gg 1$, we can utilize the semi-classical limit to approximate expectations of occupations as simply $f_\text{amp}$. Since $\Delta n^2(t)$ is a square of a quadratic observable, we expect it to scale as $\sim {f_\text{amp}}^2$ for both limits. For the other two observables ${\mathcal{G}}(x,t)$ and ${\mathcal{C}}(x,t)$, however, we must exercise care as they contain commutators. We can make predictions about these observables in the semi-classical limit $f_\text{amp} \gg 1$. Generally, the commutator `eats up' operators, leaving a commutator of two operators of as the same order as the commutands themselves
So, we can approximate the commutator as a random variable of order $f_\text{amp}$. Since the mean vanishes, we expect it to be distributed $\sim \pm f_\text{amp}$, which means that the absolute value has the expectation value $\sqrt{f_\text{amp}}$ for each disorder realization. Since ${\mathcal{G}}(x,t)$ is the square of the absolute value of the commutator, we predict a linear scaling ${\mathcal{G}}_{\max} \sim f_\text{amp}$. On the other hand, the OTOC measures the square of the commutator, and therefore has the expectation value ${f_\text{amp}}^2$ for each disorder realization. Therefore, we can predict a ${\mathcal{C}}_{\max} \sim {f_\text{amp}}^2$ for $f_\text{amp} \gg 1$.
In Fig. \ref{fig:amps_slopes} (a), we show the scalings of $\mathcal{O}_{\max}$ for the three studied observables. We see the scaling for ${\Delta n^2}_\text{max}$ is very close to $\sim {f_\text{amp}}^2$, as predicted by the analytic arguments. ${\mathcal{G}}_\text{max}$ is also consistent with a linear scaling, $\sim {f_\text{amp}}$. We note that surprisingly, ths scaling holds for $f_\text{amp} < 1$ as well. Lastly, we see that ${\mathcal{C}}_\text{max}$ appears to be consistent with interpolation between linear and quadratic behaviour in $f_{\text{amp}}$.

Next, we study how the power-law decay constant $\alpha$ changes with occupation amplitude $f_\text{amp}$. To come up with an analytic argument, we choose a similar state as Eq. \eqref{eqn:01trace}, but instead of a trace up to one particle, we take a trace up to $f_\text{amp}$ particles. For this state, instead of the cosine factor that we see in Eq. \eqref{eqn:timeflucapprox}, which arises from summing $e^{-itF^{lm}_k n}$ for $n=0,1$, we instead get a truncated geometric series for $n=0, \dots, f_\text{amp}$. The magnitude of the equivalent factor is $\sim \lvert \sin((f_\text{amp}+1) F^{lm}_k t/2) / \sin(F^{lm}_k t/2) \rvert$, whose maximum value scales as $\sim f_\text{amp}$. Therefore, we can approximate factors inside of the $d$-ball as a constant proportional to $f_\text{amp}$, 
$C f_\text{amp}$. This results in the time-dependent power law to scale as $\alpha \sim \ln(C f_\text{amp}) \, \xi_\text{loc}$. However, in the limit $f_\text{amp} \rightarrow 0$, we expect $\alpha \rightarrow 0$. This is because dynamics are caused by interactions, which is suppressed if there are no particles. This argument is numerically supported in Fig. \ref{fig:amps_slopes} (b), where we see that plot of $\exp(\alpha)$ vs. $f_\text{amp}$ results in a straight line, for both $\Delta n^2(t)$ and ${\mathcal{G}}(x,t)$.

\section{Conclusion} \label{sec:conclusion}

In this work, we have formulated a method to solve for the dynamics in 1D and 2D bosonic quantum matter in a deep MBL phase at strong disorder and weak interactions. This method represents an extension of the approach described in Ref.~\cite{detomasi19} and yields an algorithm requiring resources depending polynomially on system sizes but being independent of the targeted time. We find that this allows us to access regimes far beyond the capabilities of other methods such as exact-diagonalization or tensor networks. Using this method, we have provided evidence that many of the MBL indicators, such as the OTOC and temporal fluctuations, which were studied and well established for fermionic and spin-chain systems, also apply to bosonic MBL matter both numerically and with analytic arguments.
Our approach can also be directly applied to other nonergodic settings such as Stark MBL~\cite{Stark_MBL_2019,van_Nieuwenburg9269}.
We have checked that the resulting physical properties very much resemble the disordered bosonic system studied in this work.

While we have been concentrating mainly on few-body observables up to OTOCs, it remains an open question to solve for the dynamics of more complicated quantities involving also higher-order correlation functions such as entanglement measures. Computing these with our method appears challenging as entanglement measures for interacting theories cannot be reduced to single-particle observables in general, especially when targeting entanglement of larger subsystems. A potential route towards accessing such quantities could be to use an approach based on correlations between random measurements~\cite{vanEnk2012, Elben2018}, which has also been used recently in an experiment to measure second-order Renyi entropies~\cite{Brydges2019}.

The present work might be considered as a first step towards exploring the dynamics of bosonic MBL matter. In the future it might be interesting to increase the accuracy of the approach. This might be achieved by accounting for the dressing of the LIOMs, which currently are taken to be the Anderson orbitals, due to interactions. A crucial next step might also be to take into account the off-diagonal interaction terms. In principle, they might be integrated out using a Schrieffer-Wolff transformation at the expense of generating higher-order diagonal terms. The resulting dynamics, however, cannot be captured anymore by the presented free-boson techniques as now the few-body correlation functions cannot be evaluated by means of Wick's theorem.

In a more general context, our approach shares similarities with the recent ideas to use classical~\cite{Schmitt2018,Verdel2021} or artificial neural network wave functions~\cite{Carleo2017,Schmitt2020,Burau2020} for the solution of Schr\"odinger's equation. Here, our method can be interpreted as a perturbative construction of such networks upon explicitly designing the network structure and computing the respective weights directly without any training or optimization~\cite{detomasi19}. Pushing our approach further to include for instance the off-diagonal interaction terms might be done by taking this connection to classical or artificial neural network wave functions and to use a time-dependent variational principle to optimize our network parameters by means of stochastic Monte-Carlo techniques. In the end it might be possible to also get closer to the experimental regimes of larger interactions and potentially towards explaining or developing a theoretical description of the pioneering experiment on bosonic MBL matter~\cite{Choi1547}.

\section*{Acknowledgement}
We would like to thank M. Knap for enlightening discussions. This project has received funding from the European Research Council (ERC) under the European Union’s Horizon 2020 research and innovation programme (grant agreement No. 853443), and M. H. further acknowledges support by the Deutsche Forschungsgemeinschaft (DFG) via the Gottfried Wilhelm Leibniz Prize program.

\appendix* 

\section{Poincar\'e-Lindstedt perturbation theory} \label{apn:pl}

In the non-interacting limit $U=0$, the Hamiltonian of Eq. \eqref{eqn:andersonbasishamiltonian} is nothing but a collection of decoupled harmonic oscillators. The presence of interactions introduces a coupling between these oscillators in a non-linear fashion. In our approximation in Eq. \eqref{eqn:lbithamiltonian}, we only consider the diagonal contribution of the interaction. This takes into account the frequency shift  of the oscillators and therefore resembles Poincar\'e-Lindstedt perturbation theory, which is a powerful perturbative approach for solving harmonic oscillators in classical systems.

Let us therefore take the chance to illustrate the working principle of our approximation using this analogy. For concreteness, let's consider the so-called Duffing equation, the simplest non-linear perturbation to the simple harmonic oscillator:
\begin{equation}
    \frac{d^2 x}{dt^2} + x + \epsilon x^3 = 0,
\end{equation}
where $0 < \epsilon \ll 1$ accounts for a weak nonlinearity. 
The Poincar\'e-Lindsted method is a version of perturbation theory, where, by considering the shift in the frequency of the unperturbed oscillatory solution, secular terms are systematically removed, resulting in a oscillatory solution that is accurate for all times. 

In conventional perturbation theory, we consider corrections in amplitude
\begin{equation}
    x(t) = x_0(t) + \epsilon x_1(t) + \cdots,     
\end{equation}
and equate each order of $\epsilon$.
For the case of the Duffing equation, the conventional perturbative solution is, to first order:
\begin{equation}
x_\text{pert}(t) = \cos(t) + \epsilon \left( \frac{1}{32} \cos(3t) - \frac{1}{32} \cos(t) - \frac{3}{8} t \sin(t) \right) \, .   
\end{equation}
We see that from the first order in $\epsilon$, a secular term  $\sim t \sin(t)$ arises, which is not bounded in time.

\begin{figure}[ht] 
\includegraphics[width=1.0\columnwidth]{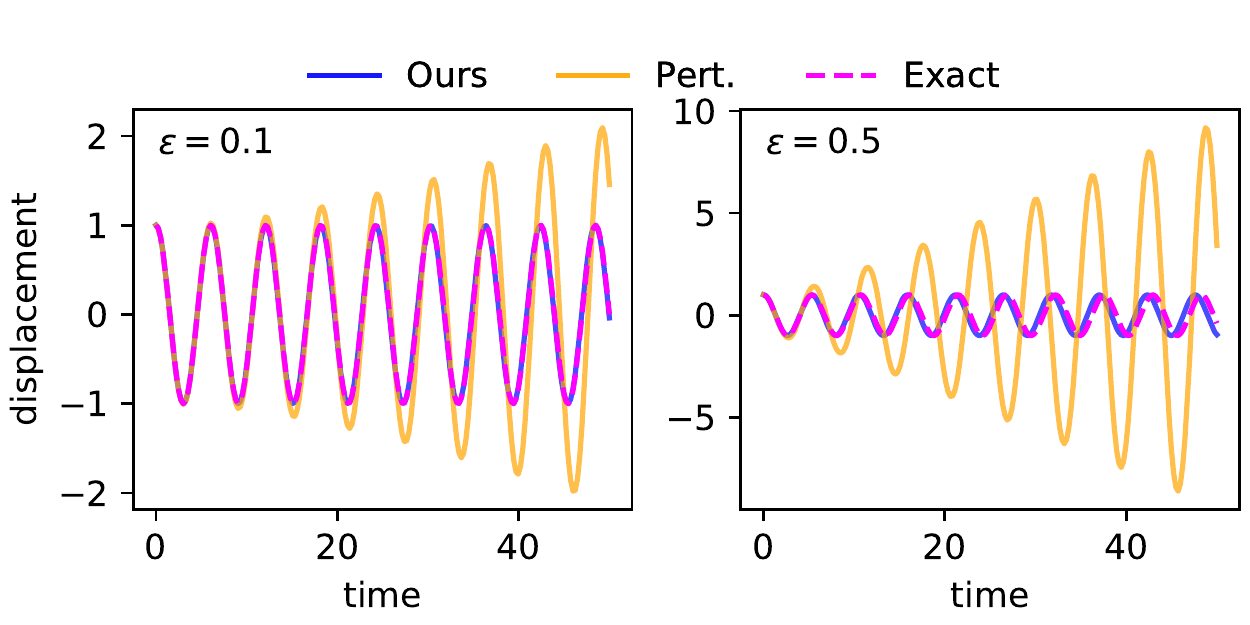}
\caption{\label{fig:duffing} Plots comparing Our method, a simplified Poincar\'e-Lindstedt perturbation theory (Ours, blue lined) with conventional perturbation theory (Pert., orange lined), to the exact solution (Exact, magenta dashed line) for the Duffing equation. Conventional perturbation theory contains secular terms that grow without bound in time.}
\end{figure}

In Poincar\'e-Lindstedt perturbation theory, on the other hand, in addition to the amplitude, we also consider the shift in frequency, as $\tau = \omega t$, where $\omega = 1 + \epsilon \omega_1 + \cdots$. By appropriate choice of $\omega_1$, the term proportional to $t \sin (t)$ can be removed, resulting in the solution 
\begin{equation}
\begin{aligned}
x_\text{PL}(t) & = \cos\left(\left[1+\frac{3}{8}\epsilon\right]t\right) \\
& + \frac{\epsilon}{32} \left[ \cos\left(3\left[1+\frac{3}{8}\epsilon\right]t\right) - \cos\left(\left[1+\frac{3}{8}\epsilon\right]t\right) \right]
\end{aligned}
\end{equation}
for the case of the Duffing equation. Within our approach we go one step further. We only consider frequency shifts and neglect any other perturbative contribution. For the above Duffing oscillator this amounts to neglecting $x_1(t)$ (and higher-order corrections), resulting in the approximation
\begin{equation}
    x_\text{Ours}(t) = \cos \left(\left[1+\frac{3}{8}\epsilon\right]t\right).
\end{equation} In Fig. \ref{fig:duffing}, we compare $x_\text{Ours}(t)$, a simplification of the Poincar\'e-Lindstedt perturbation theory (blue line), with the conventional perturbation theory (orange line) and the exact solution (magenta dashed line). We see that even though Our solution neglects the term proportional to $\epsilon$ in the Poincar\'e-Lindstedt solution, it is still accurate for moderate perturbative strength $\epsilon$ and stays bounded over time, whereas the conventional perturbation theory breaks down for times $t \gtrsim 1/\epsilon$ due to the presence of secular terms.

\bibliography{apssamp}

\end{document}